\documentclass[prl,twocolumn,amsfonts,amsmath,tightenlines,preprintnumbers,nofootinbib,letterpaper]{revtex4}

\usepackage{amsmath,amssymb}
\usepackage{graphicx}
\usepackage{bm}
\usepackage{comment}

\newcommand{\beq}{\begin{equation}}
\newcommand{\eeq}{\end{equation}}
\newcommand{\bea}{\begin{eqnarray}}
\newcommand{\eea}{\end{eqnarray}}

\def\OMIT#1{{}}

\newcommand{\lsim}{\ \raisebox{-0.7ex}{$\stackrel{\textstyle <}{\sim}$ }}

\newcount\hour \newcount\hourminute \newcount\minute 
\hour=\time \divide \hour by 60
\hourminute=\hour \multiply \hourminute by 60
\minute=\time \advance \minute by -\hourminute

\begin{document}
\begin{figure}[!t]
  \vskip -1.5cm
  \leftline{\includegraphics[width=0.15\textwidth]{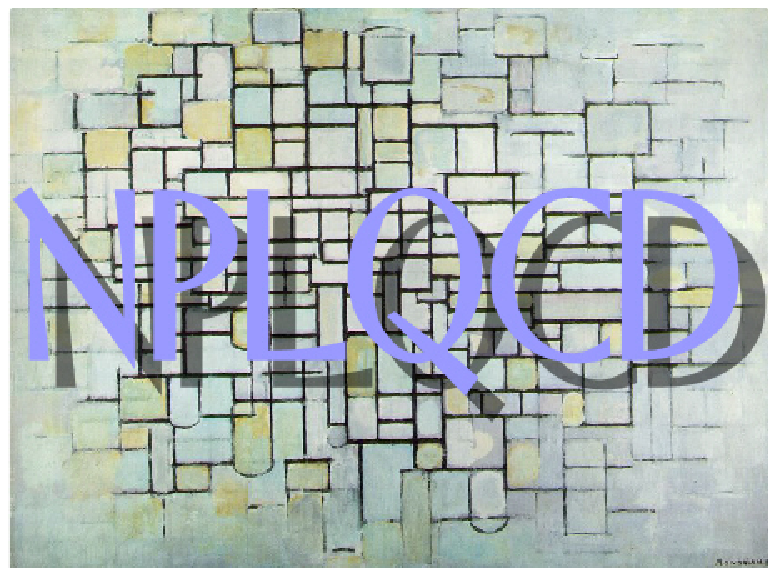}}
\end{figure}

\preprint{\vbox{
    \hbox{NT@UW-08-17\ \ ,\ \ Athena-pub-08-01}
  }}

\title{Color Screening by Pions}
\author{William Detmold and Martin J. Savage}
\affiliation{
Department of Physics, University of Washington, 
Seattle, WA 98195-1560.\\
(NPLQCD Collaboration)
}
\date{\today}

\vskip 0.8cm
\begin{abstract}
\noindent 
Lattice QCD is used to calculate the potential between a static quark and anti-quark
in the presence of a finite density of $\pi^+$'s.
Correlation functions of multiple $\pi^+$'s
are used in conjunction with Wilson-loop correlators 
to determine the difference between the $Q\overline{Q}$-potential  in
free-space and in the presence of a pion condensate.  
The modifications to the potential are found to have significant dependence on
the $Q\overline{Q}$ separation over the range $r\lsim 1~{\rm fm}$
explored in this work.
Our results are consistent with the pion-condensate behaving as
a (nonlinear) chromo-dielectric.
\end{abstract}
 \maketitle

\vfill\eject

%
%

\noindent {\it Introduction:}\ 
Flavor dynamics plays a central role in the study of heavy-ion collisions and,
ultimately, in unraveling the phase structure of matter at finite temperature and
density.
The production, evolution and detection of the heavy flavors, charm and bottom,
in such collisions has long been known to be sensitive to the presence of 
deconfined phases of Quantum Chromodynamics (QCD), such as 
the naive quark-gluon plasma or  ``the perfect fluid'' seen at RHIC~\cite{Adams:2004bi,Adare:2006nq}.
Color charges are screened in a deconfined phase, and 
the production of a given state of quarkonium is expected 
to be suppressed when the screening length
becomes comparable to, or smaller than, its size~\cite{Matsui:1986dk}
 (for a
recent discussion see Ref.~\cite{Satz:2006kba}).
However, quantitative calculations of ``$J/\psi$-suppression'' and more generally ``quarkonium-suppression''
in heavy-ion collisions are difficult due to the enormous complexity
of such collisions.

Understanding the behavior of a  static quark, $Q$, and
static anti-quark, $\overline{Q}$,  pair produced in a
heavy-ion collision requires knowledge of its dynamics in both the deconfined
and hadronic phases.
In the hadronic phase, the potential 
between a static quark and static anti-quark (denoted as  the $Q\overline{Q}$ potential)
can be screened by the
hadrons that participate in the collision.
Differences between this ``hadronic medium effect'' and the effect of a
deconfined phase are  tell-tale signatures of a deconfining phase transition 
in heavy-ion collisions.
As a step to improving the description of $Q\overline{Q}$ transport in
the hadronic phase
we present the results of a Lattice QCD
calculation of the modifications of the 
$Q\overline{Q}$
potential and force 
resulting from  the presence of a condensate of
charged pions (either all $\pi^+$'s or all $\pi^-$'s)
in the absence of electromagnetism.  
The $Q\overline{Q}$-potential is found to be lowered
in the condensate, with a non-trivial dependence upon the $Q\overline{Q}$
separation, 
leading to a reduced $Q\overline{Q}$ force.

It is useful to consider the behavior of the in-medium $Q\overline{Q}$ potential and force
in the limiting cases where the $Q\overline{Q}$ separation, $R$, is much smaller
than, or much greater than, the scale of chiral symmetry breaking,
$\Lambda_\chi$, that typifies hadronic interactions.
In the   $R \ \Lambda_\chi \ll 1$ limit,
it is appropriate
to construct an effective field theory (EFT) describing the interactions
between the 
$Q\overline{Q}$ with the gluon fields.  The coefficients in this EFT are
determined via a multi-pole expansion of the matrix elements calculated in QCD.
The interactions between the $Q\overline{Q}$-state and the
hadronic background are thereby factorized into short-distance parts
encapsulated in the coefficients of 
the EFT, and low-energy matrix elements of local operators composed of quark and
gluon fields (explicitly independent of $R$).  
This method~\cite{Peskin:1979va} has been used to calculate the binding energy of
quarkonium to infinite nuclear matter~\cite{Luke:1992tm}, and 
quarkonium scattering from nucleons~\cite{Brodsky:1997gh}.
The interactions of a spatially-averaged and orientation-averaged 
Wilson-loop with quark and gluon fields, at leading order
in the strong-coupling and derivative expansion, 
are described by  an effective  Lagrange density of the form
\begin{eqnarray}
{\cal L} & = & 
R^3\ 
{\cal S}^\dagger {\cal S}\ 
G_{\mu\alpha} G_{\nu}^\alpha
\left[\ 
c_1 g^{\mu\nu}
\ +\ 
c_2 v^\mu v^\nu
\ \right]
\ ,
\label{eq:smallR}
\end{eqnarray}
where ${\cal S}$ is the operator that annihilates the $Q\overline{Q}$-pair,
$G_{\alpha\beta}$ is the gluon field strength tensor, 
$v_\alpha$ is the $Q\overline{Q}$-pair four-velocity.
The
renormalization-scale dependent coefficients
$c_i$ that appear in eq.~(\ref{eq:smallR}) are dimensionless and are expected
to be of order unity by naive dimensional analysis.
The Lagrange density in eq.~(\ref{eq:smallR}) makes explicit the 
separation dependence of the in-medium component of the potential ($\propto R^3$)
and force ($\propto R^2$) for  $R \ \Lambda_\chi \ll 1$.

At extremely large distances, the ground state of the system is a
heavy meson--anti-meson pair in the pion condensate
interacting through the exchange of hadrons
described by an effective field theory analogous to that used to describe the
interactions of nucleons.
In QCD, the force between these heavy mesons is Yukawa-like  with a mass-scale
set by the pion mass.

At intermediate distances and in vacuum, the ground-state of the system has the 
$Q\overline{Q}$ potential increasing with separation, and the gluonic field
configuration between the $Q\overline{Q}$ is tending toward a flux-tube 
with constant force between the $Q\overline{Q}$ pair 
(for recent lattice calculations see Ref.~\cite{Bali:2005fu}). 
It is not obvious how the presence of a pion-condensate will modify the 
interactions between the $Q\overline{Q}$ pair.  
If instead, the system under consideration was a collection of neutral
hadrons in the presence of an electric field,
the energy-shift of the system would depend 
upon the electric-polarizabilities of the hadron and the volume of the electric
field.  An overall reduction in the $Q\overline{Q}$ force would result,
encapsulated in the dielectric function of the medium.

\noindent {\it The Lattice QCD Calculation:} 
The ground-state energy of a system comprised of a
$Q\overline{Q}$-pair separated by a distance $R$
and $n$-$\pi^+$'s can be extracted from the correlation functions
\begin{eqnarray}
  \label{eq:corrs}
&&  C_n(t_\pi,t) = \Big\langle 0 \Big| \Big[\sum_{\bf x} \chi_{\pi^+}({\bf
      x},t) \chi^\dagger_{\pi^+}(0,t_\pi)\Big]^n \Big|0\Big\rangle
\ \ ,
\nonumber\\
&&C_W (R,t_w,t) =   \Big\langle 0 \Big| 
\sum_{{\bf y},|{\bf r}|=R}
{\cal W}\left({\bf y}+{\bf r}, t;{\bf y},t_w\right)
\Big|0\Big\rangle
\ \ ,
\nonumber\\
&&C_{n,W} (R,t_\pi,t_w,t) =   \Big\langle 0 \Big| 
\Big[\sum_{\bf x} \chi_{\pi^+}({\bf x},t) \chi^\dagger_{\pi^+}(0,t_\pi)\Big]^n
\nonumber\\
&& \qquad  \qquad
\times
\sum_{{\bf y},|{\bf r}|=R}
{\cal W}\left({\bf y}+{\bf r}, t; {\bf y},t_w\right)
\Big|0\Big\rangle
\ \ ,
\end{eqnarray}
where $\chi_{\pi^+}(x)=u^a (x) \gamma_5 \overline{d}_a(x)$ is a
pseudoscalar interpolating operator for the $\pi^+$ ($a$ is a color
index), and ${\cal W}\left({\bf y},t_0; {\bf y}+{\bf r}, t\right)$ is the
Wilson-loop operator formed from products of gauge-links 
joining the vertices at $({\bf y},t_0)$, $({\bf y}+{\bf r},t_0)$,
$({\bf y}+{\bf r},t)$ and $({\bf y},t)$.
At large times, the static $Q\overline{Q}$ potential in vacuum is extracted
using
\begin{eqnarray}
  \sum_{t_w}\ C_W(R,t_w,t_w+t)  & \stackrel{t\to\infty}{\longrightarrow} &
A \ e^{-V(R)\  t}
\ \ .
  \label{eq:vacrat}
\end{eqnarray}
It is useful to define
the ratio of the three correlation functions in eq.~(\ref{eq:corrs}),
\begin{eqnarray}
G_{n,W}(R,t_\pi,t_w,t) & = & 
{ C_{n,W} (R,t_\pi,t_w,t)\over C_n(t_\pi,t) \ C_W (R,t_w,t)
}
\ \ ,
\label{eq:medrat}
\end{eqnarray}
from which the in-medium modification to the potential can be extracted, for
$t_w\gg t_\pi$ (ensuring that the pion system is in its ground-state), 
\begin{eqnarray}
\langle\log\left( G_{n,W}(R,t_\pi,t_w,t_w+t)\over  G_{n,W}(R,t_\pi,t_w,t_w+t+1)
\right)
\rangle_{t_w}
 \stackrel{t\to\infty}{\longrightarrow} 
\delta V(R,n)
\ , &&
\label{eq:medpot}
\end{eqnarray}
where the $\langle...\rangle_{t_w}$ denotes an average over a number of initial
time-slices for the Wilson-loop, $t_w^{\rm min}$ to $t_w^{\rm max}$.

We have computed the correlators in eq.~(\ref{eq:corrs})
in mixed-action lattice QCD, using
domain-wall valence quark propagators from a Gaussian smeared-source
on rooted-staggered MILC gauge-configurations
(see Refs.~\cite{Aubin:2004fs,Beane:2007xs} for details). 
Here we focus on calculations on an ensemble of 1001 lattices with 
a pion mass of $m_\pi\sim~320~{\rm MeV}$, and a
lattice spacing of $b=0.087(1)~{\rm fm}$ with dimension $28^3\times 96$ giving 
a spatial dimension of $\sim~2.5~{\rm  fm}$.  
Propagators were calculated with both periodic and anti-periodic boundary
conditions in the time-direction
and combined  to effectively double the
length of the time-direction leading to long plateaus in effective energy
plots of the mesonic correlators.
The light quark propagators were computed after the gauge-field had undergone a
single level of HYP-smearing~\cite{Hasenfratz:2001hp}, while Wilson-loops were calculated with $N_{\rm
  HYP}=0,1,2$ and $4$ levels of HYP-smearing of the gauge-field.  
Further, the spatial links of the Wilson-loops  were 
APE-smeared~\cite{Teper:1987wt,Albanese:1987ds}
 in the
transverse-spatial directions
to optimize the signals for $C_W (R,t_w,t)$.
The increasing levels of  HYP-smearing 
result in improved  signal-to-noise ratios and 
enable the
potential, and in-medium modifications to the potential, to be determined over a
range of $Q\overline{Q}$ separations.
The correlation functions $C_n(t_\pi,t)$ have been previously calculated on 
these lattices, enabling a
study of the properties of the pion and kaon condensates, and the three-meson
interactions~\cite{Beane:2007es,Detmold:2008fn,Detmold:2008yn}.

\begin{figure}[!t]
  \centering
\includegraphics[width=3.0in]{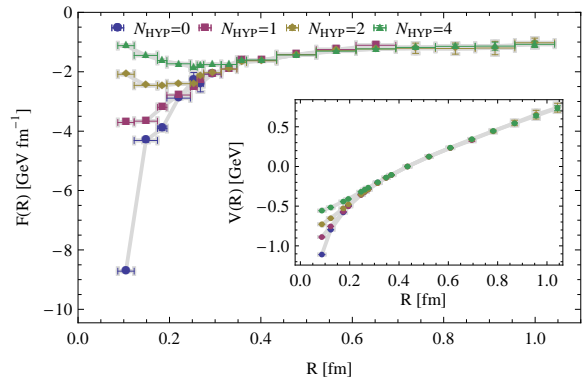}  
  \caption{The $Q\overline{Q}$ force in vacuum determined with different levels
    of 
    HYP-smearing.  Inset is the $Q\overline{Q}$ potential with these same
 smearings, each normalized to $V(R_0)=0$ at  $R_0=5 b$.
} 
  \label{fig:Fhyp}
\end{figure}
Correlated fits are performed
to  the effective energies (or energy-differences),
{\it e.g.} eq.~(\ref{eq:medpot}), for each 
$Q\overline{Q}$ separation and number of pions in the volume. 
Separately, jackknife and bootstrap procedures were used to generate the covariance matrix
over the given fitting interval of time-slices, and correlated
$\chi^2$-minimization was performed to extract the energy and its associated
statistical uncertainty.
A systematic uncertainty is determined by a comparison of the
various fit procedures and various fitting ranges (including the choice of
$t_w^{\rm min/max}$).

\begin{figure}[!t]
  \centering
\includegraphics[width=3.1in]{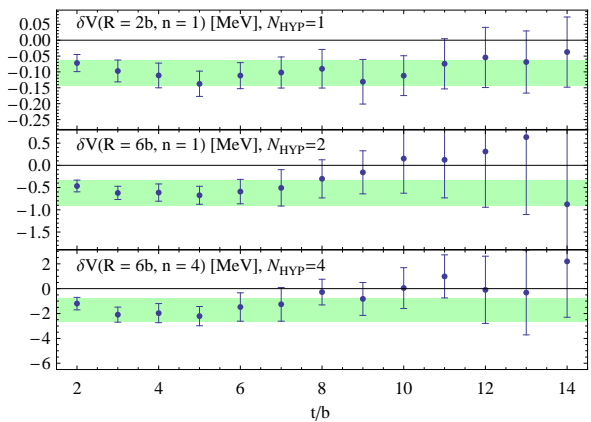}  
  \caption{Representative effective energy plots for the in-medium
    contributions to the potential. 
The horizontal band corresponds to the combined statistical and systematic
uncertainty of the respective fit.
} 
  \label{fig:Effenergyplots}
\end{figure}

The $Q\overline{Q}$-potential in vacuum is determined from $C_W$, and the 
$Q\overline{Q}$-force is determined by finite-differences
of the potential.
Our  results for these two quantities are shown in fig.~\ref{fig:Fhyp}.
By comparing the force in vacuum calculated with different levels of
HYP-smearing, we ascertain the separations at which the force (and potential)
cease to be significantly contaminated by the smearing procedure.
We conclude from
fig.~\ref{fig:Fhyp} that the potential and force calculated at
separations $R >  b N_{\rm HYP}$ are close to the result of an un-smeared
calculation.  
In our analysis, we only use calculations that satisfy this criterion.
Three representative effective energy plots 
associated with the in-medium contributions to the potential
are shown in fig.~\ref{fig:Effenergyplots}.
Typically, signals become noisier as either $R$ or $n$ increases, restricting
the present  analysis to $R\lsim 1~{\rm fm}$.

\begin{figure}[!b]
  \centering
\includegraphics[width=3.1in]{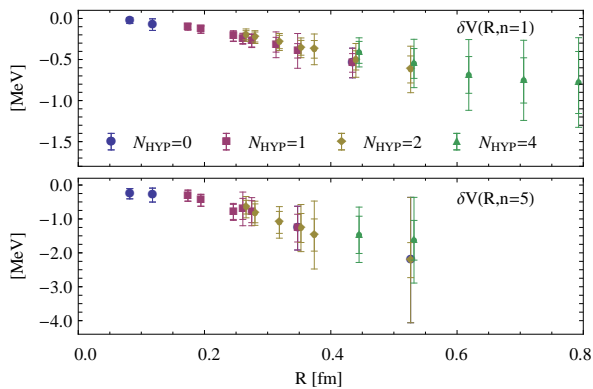}  
  \caption{The in-medium contributions to the 
$Q\overline{Q}$ potential at a pion densities of $\rho_0$ and $5\rho_0$.
The inner uncertainty associated with each point
is statistical, while the outer is the statistical and
systematic uncertainties combined in quadrature.
Different HYP-smearings are offset for clarity.
} 
  \label{fig:dVhyp}
\end{figure}

\begin{figure}[!t]
  \centering
\includegraphics[width=3.1in]{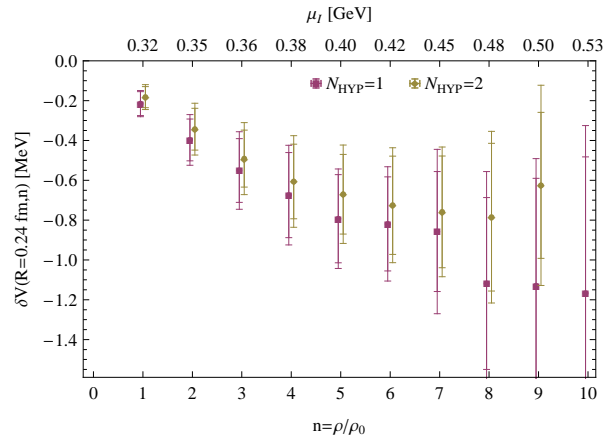}  
  \caption{The in-medium contribution to the $Q\overline{Q}$
potential at $R\sim 0.24~{\rm fm}$
as a function of the number of pions.
The uncertainties are as described in fig.~\protect\ref{fig:dVhyp}.
The isospin chemical potential of the system 
is shown on the upper axis~\protect\cite{Detmold:2008yn}.
} 
  \label{fig:dVnpi}
\end{figure}

The in-medium contributions to the $Q\overline{Q}$-potential, $\delta V(R,n)$,
resulting from both 
a single pion in the lattice volume 
(a number density of $\rho_0\sim 1/(2.5~{\rm fm})^3=0.064~{\rm fm}^{-3}$), 
and from five pions in the lattice volume ($\rho=5\rho_0=0.32~{\rm fm}^{-3}$),
are shown in 
fig.~\ref{fig:dVhyp}.
For $\rho\lsim 7\rho_0$ they are found to be linear in the density 
within the uncertainties of the calculations, as can be seen 
for a representative $R$  in fig.~\ref{fig:dVnpi}.
We note that in-medium contributions to the potential as small
as $\delta V(b,1)\sim 100~{\rm keV}$ have been determined.
For the system containing a single pion in the lattice volume, the energy-shift
can be directly related to the scattering phase-shift using L\"uscher's method~\cite{Luscher:1986pf}.
Therefore, the $\delta V(R,1)$ shown in fig.~\ref{fig:dVhyp} can be used to
determine the scattering length associated with a $Q\overline{Q}$-pair of
fixed separation and a pion.

\begin{figure}[!b]
  \centering
\includegraphics[width=3.1in]{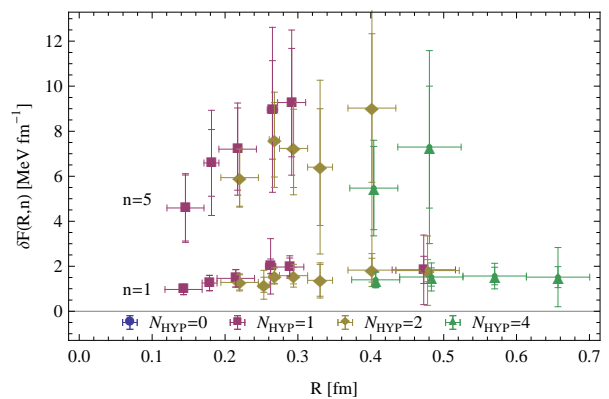}  
  \caption{The in-medium contribution to the radial $Q\overline{Q}$ force, $\delta F(R,n)$,
at a pion number density of $\rho_0$ and $5\rho_0$.
The uncertainties are as described in the caption of fig.~\protect\ref{fig:dVhyp}.
} 
  \label{fig:dF}
\end{figure}

The effects of the medium on  the radial $Q\overline{Q}$ force, $\delta F(R,n)$, 
are determined from the effective energy-differences derived from 
finite differences of $\delta V(R,n)$ with respect to $R$.
The modifications to the force at densities  $\rho_0$ and $5\rho_0$
are shown in fig.~\ref{fig:dF}.
The  $Q\overline{Q}$ force is seen to be reduced by an approximately
$R$-independent amount over the separations 
and pion densities we have been able to explore.

The in-medium potential and force have also been calculated on
MILC lattices of the same spatial volume as the configurations used here,
at four different quark masses, but at a coarser lattice spacing ($b\sim 0.125~{\rm fm}$).
We find the results at the corresponding pion mass are consistent, 
suggesting that lattice discretization errors are not large.
However,  the uncertainties in the medium
modifications to the $Q\overline{Q}$
potential and force are somewhat larger than those on the current
ensemble.
A mild variation with  the pion mass was observed, however,  more precise
calculations are required in order to quantify this dependence.

\noindent {\it Discussion:}
In-medium effects play an important  role in the diagnostics used to explore new
phases of matter in heavy ion collisions, and more generally emerge as useful
quantities in the context of mean-field constructions in many-body systems.
We have performed the first QCD calculation (without electromagnetism)
of modifications to
the $Q\overline{Q}$ potential and force in a hadronic medium
by calculating the energy of a Wilson-loop in the presence of a condensate of
charged pions.
The attractive $Q\overline{Q}$ interaction is found to be reduced by the
hadronic medium
over the range of separations we were able to explore, $R\lsim 1~{\rm fm}$.
This is a first step
toward a more systematic exploration of in-medium effects with Lattice QCD,
with the ultimate goal of looking for in-medium modifications of hadronic
observables in backgrounds of baryons.
Such calculations will require
significantly more computational resources than are currently available,
and precise calculations of multi-baryon systems are a prerequisite.

At $\rho_0$, our calculations
are analogous to those 
of $J/\psi$-$\pi$ scattering lengths
(with the $J/\psi$ replaced by a Wilson-loop)
which have been performed  previously in 
QCD~\cite{LiuPOSTER} and 
quenched QCD~\cite{Yokokawa:2006td}. 
At higher densities, the calculations involve multi-pion backgrounds and are the
first of their kind.
A  non-zero three-pion interaction is required to
describe the volume-dependence of the energy-levels of $n>2$ $\pi^+$'s in these
volumes~\cite{Beane:2007es,Detmold:2008fn,Detmold:2008yn}, 
however,  within the uncertainties of our
calculations,
the multi-pion interactions with the
$Q\overline{Q}$-pair are found to be consistent with zero over the range of
$Q\overline{Q}$ separations we have explored.  
It is important to refine this work by performing 
higher-statistics calculations in order to determine the
multi-pion interactions with the $Q\overline{Q}$-pair.
In addition, with a corresponding calculation of the pionic matrix elements of
the gluonic operators in eq.~(\ref{eq:smallR}), the coefficients, $c_i$, in 
eq.~(\ref{eq:smallR}) could be determined.

{}From our analysis we observe
that the medium-modification of the force is independent of the $Q\overline{Q}$
separation in the region where the force in vacuum is becoming constant.
This 
is consistent with the pion-condensate behaving as a dielectric
in the volume of the color flux-tube between the $Q\overline{Q}$-pair.
It implies that the pion and, collectively, the condensate 
has a chromo-susceptibility,
which is expected to be highly non-linear in the
gluon-field strength.
At small separations, the modification to the force appears  consistent with the
behavior expected from eq.~(\ref{eq:smallR}),
but calculations at smaller lattice spacing are required to confirm this.

\vskip 0.2in
\acknowledgments{
We thank our NPLQCD collaborators for their contributions to this work,
Barak Bringoltz, David Kaplan and Jerry Miller for useful conversations, and R.~Edwards and B.~Joo for help with the
QDP++/Chroma programming environment~\cite{Edwards:2004sx}.  
We are indebted to
the MILC for use of their configurations.  
This work was supported in part by the U.S.~Dept.~of Energy under Grant No.~DE-FG03-97ER4014. 
The
computations for 
this work were performed at Jefferson Lab, Fermilab, Centro Nacional de
Supercomputaci\'on, 
the University of Washington (the Athena cluster),
NERSC (Office of
Science of the U.S. Department of Energy,
No. DE-AC02-05CH11231), and the
NSF through TeraGrid resources provided by the
NCSA. 
}

%
%

\end{document}